\documentclass[9pt,twocolumn,twoside]{opticajnl}
\journal{opticajournal} % use for journal or Optica Open submissions
\setboolean{shortarticle}{true}
\usepackage{setspace}
\usepackage{lineno}
\usepackage{graphicx}
\usepackage{color,soul}

\usepackage{float}

\title{Three-dimensional, multi-wavelength beam formation with integrated metasurface optics for Sr laser cooling}

\author[1,2]{Sindhu Jammi}
\author[1,2]{Andrew R. Ferdinand}
\author[1,2]{Zheng Luo}
\author[4]{Zachary L. Newman}
\author[1,2]{Gregory Spektor}
\author[3]{Junyeob Song}
\author[3]{Okan Koksal}
\author[5]{Akash V. Rakholia}
\author[5]{William Lunden}
\author[5]{Daniel Sheredy}
\author[5]{Parth B. Patel}
\author[5]{Martin M. Boyd}
\author[3]{Wenqi Zhu}
\author[3]{Amit Agrawal}
\author[1]{Travis C. Briles}
\author[1,2,*]{Scott B. Papp}

\affil[1]{Time and Frequency Division, National Institute of Standards and Technology, Boulder, Colorado, 80305, USA}
\affil[2]{Department of Physics, University of Colorado, Boulder, Colorado, 80309, USA}
\affil[3]{Microsystems and Nanotechnology Division, National Institute of Standards and Technology, Gaithersburg, Maryland, 20899, USA}
\affil[4]{Octave Photonics, Louisville, Colorado, 80027, USA}
\affil[5]{Vector Atomic, Inc., Pleasanton, California 94588, USA}

\affil[*]{scott.papp@nist.gov}

\begin{abstract} %approx 100 words.  Soft limit
We demonstrate the formation of a complex, multi-wavelength, three-dimensional laser beam configuration with integrated metasurface optics. Our experiments support the development of a compact Sr optical-lattice clock, which leverages magneto-optical trapping on atomic transitions at 461 nm and 689 nm without bulk free-space optics. We integrate six, mm-scale metasurface optics on a fused-silica substrate and illuminate them with light from optical fibers. The metasurface optics provide full control of beam pointing, divergence, and polarization to create the laser configuration for a magneto-optical trap. We report the efficiency and integration of the three-dimensional visible laser beam configuration, demonstrating the suitability of metasurface optics for atomic laser cooling.
\end{abstract}

\setboolean{displaycopyright}{false} % Do not include copyright or licensing information in submission.

\begin{document}

\maketitle

Laser-cooled gases of alkaline-earth atoms have led to revolutionary advances in atomic clocks \cite{Ludlow2015OpticalClocks} and precision measurement \cite{Bothwell2022ResolvingSample}. The magneto-optical trap (MOT), which consists of three pairs of intersecting laser beams and a magnetic quadrupole field, has been the dominant method of preparing ultracold atomic vapors since its invention decades ago \cite{Raab1987TrappingPressure}. The appropriate configuration of optical powers and state of circular polarization results in balanced conservative and dissipative forces that spatially confine and reduce the temperature of an atomic sample.  In a research laboratory, MOTs are typically built from an array of optical components that must be carefully aligned, taking up a volume of $\sim$1 m$^3$.  This level of complexity is a significant barrier to the development of portable cold atom systems \cite{Burrow2021Stand-aloneTechnologies,Lee2022ASystem} that can address applications in geodesy \cite{McGrew2018AtomicLevel} and inertial sensing \cite{Dickerson2013MultiaxisInterferometry}.  In particular, integrated MOTs would greatly benefit the development of compact optical lattice clocks based on the ultra-narrow, spin-forbidden $^1$S$_0$ $\rightarrow$ $^3$P$_0$ transition in $^{87}$Sr and other alkaline-earth systems \cite{Ludlow2015OpticalClocks}. Recently, there has been a push towards using micro- or nanofabricated integrated photonic systems \cite{McGilligan2022Micro-fabricatedSensors} that are compatible with foundry-scale manufacturing to reduce the MOT size, complexity and cost.

\begin{figure}[ht!]
\centering
\includegraphics[width=\linewidth]{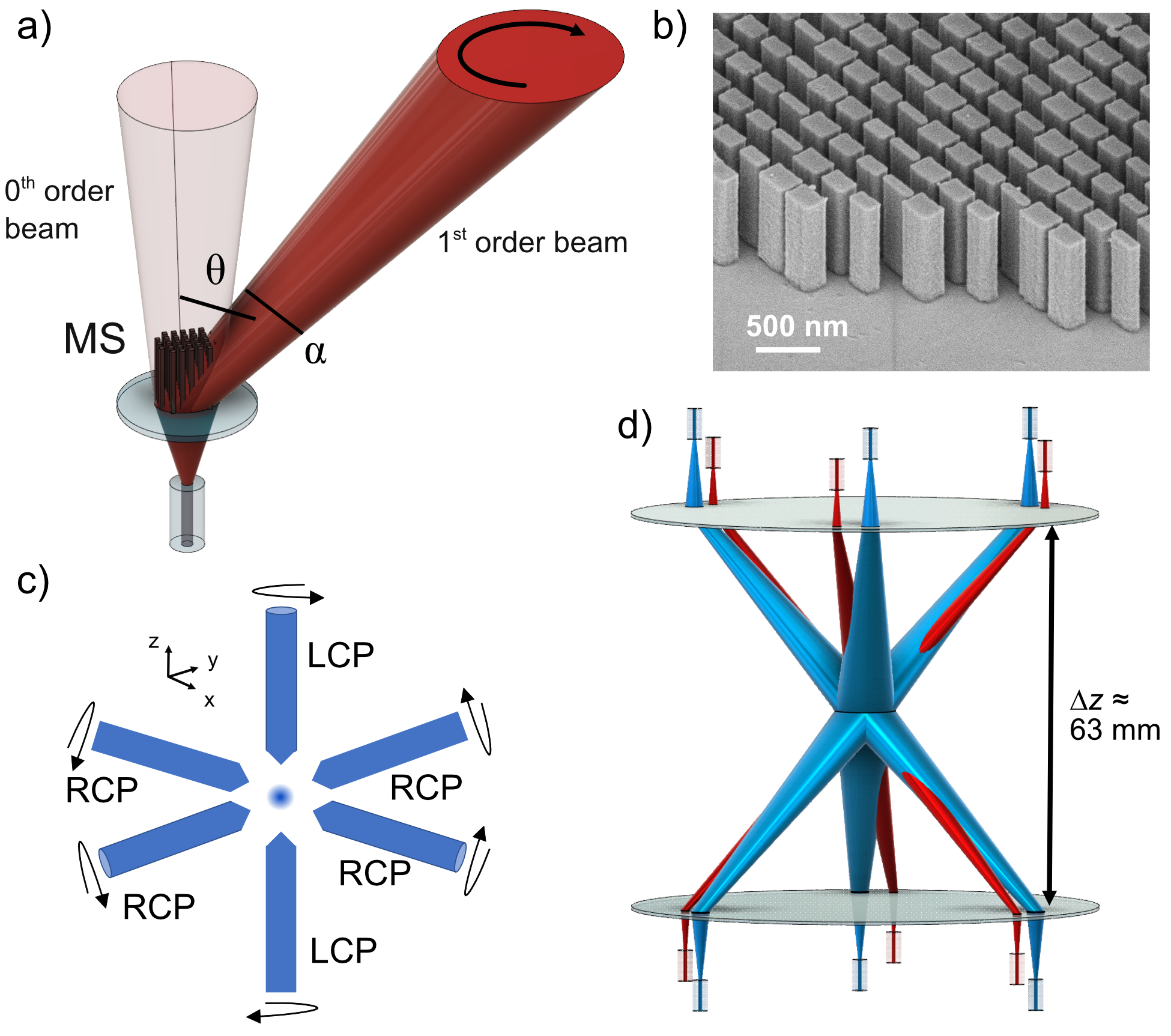}
\caption[\parbox{3.25in}]{
Metasurface (MS) functionality and their integration for two-stage Sr MOTs.   
a) Each MS controls the deflection angle $\theta$, divergence angle $\alpha$ and the circular polarization (black arrow) of the 1$^{\text{st}}$-order beam. 
b) SEM image of the MS. 
c) Beam geometry and polarization in a conventional cubic MOT. LCP (RCP) stands for left (right) circular polarization.  
d) Multi-wavelength MS integration on fiber-coupled, 3-inch wafers to produce all 12 trapping beams for 461 nm and 689 nm MOTs; wafer separation,  $\Delta z$.
}
\label{fig:fig1}
\end{figure}

Compared to alkali atoms like rubidium, strontium poses unique challenges to any photonic integration for MOTs. The broad linewidth of the $^1$S$_0$ $\rightarrow$ $^1$P$_1$ cooling transition at 461 nm and the lack of sub-Doppler cooling mechanisms requires a second stage MOT on the narrow $^1$S$_0$ $\rightarrow$ $^3$P$_1$ transition at 689 nm to reach $\mu$K temperatures. Previous work on integrated Sr MOTs with pyramid reflectors has suffered from limited optical access to the atom cloud \cite{Bowden2019AClock}. The planarized geometry of grating MOTs \cite{Sitaram2020ConfinementTrap} addresses this problem, but such systems are unable to achieve the level of precise beam control found in traditional MOTs built from bulk optics.  This limitation is problematic when cooling atoms with complex level structures such as $^{87}$Sr \cite{Barker2023GratingStructures}. An alternative integration approach is to use metasurface (MS) optics which consist of periodic arrays of dielectric nanopillars \cite{Arbabi2022AdvancesMetalenses}.  These devices are capable of flexible control of the optical phase and offer the potential to perform the function of multiple traditional optics simultaneously in a single wafer-thickness optic. To date, atom trapping experiments with MSs have underutilized their multifunctional capabilities and have required auxiliary optical components to achieve the necessary level of control. Furthermore, the demonstrated MS beam delivery methods represent significant challenges to creating scalable systems.  Free space coupled MS systems \cite{Zhu2020AAtoms,McGehee2021Magneto-opticalOptics,Hsu2022Single-AtomTweezer} still require optical alignment and assembly  while guided mode approaches using photonic integrated circuits (PICs) \cite{Yulaev2019Metasurface-IntegratedControl, Spektor2021InverseClock,Isichenko2023PhotonicTrap} exhibit high propagation loss of distances of $\approx 10$ mm, especially at the visible wavelengths necessary to trap Sr \cite{Ropp2023IntegratingChip}.

Here, we demonstrate the integration of metasurface optics on a common substrate to generate the laser-beam configuration for Sr laser cooling at 461 nm and 689 nm. Our metasurface optics platform enables the generation of the complete beam configuration of a MOT directly from light emitted by optical fibers without any bulk free-space optics.   Additionally, our foundry-compatible integration strategy avoids the high-propagation loss found in PIC approaches and eliminates the need for assembly and optical alignment required for traditional MOTs. Instead, these features are automatically incorporated into our design at the nanophotonic-level by integrating multiple functionalities into a single MS optic, and at the system level by integrating MS designs for different beams and wavelengths on semiconductor wafers. Each MS simultaneously performs the beam formation functions of conventional steering mirrors and lenses, as well as the polarization transformation of quarter waveplates (QWPs). We present a comprehensive characterization of the design and performance of our nanofabricated structures, underscoring their role as fundamental building blocks in integrated, nanophotonic optical systems for diverse applications in cold-atom quantum computing and quantum sensing.

Figure \ref{fig:fig1} introduces our metasurface optics devices and describes how we use them as compact replacements for traditional MOT optics.  The basis for beam control with MSs is the spatially dependent phase shift that results from the interaction of the optical beam with nanopillars of specific sizes placed across the transverse $x$-$y$ plane \cite{Arbabi2022AdvancesMetalenses}. We set the size and distribution of the nanopillars to design a polarization-dependent MS phase function that transforms the linearly polarized input optical mode from an optical fiber into a $1^{\text{st}}$-order beam deflected by angle $\theta$ with efficiency $\eta$, focused to full divergence angle $\alpha$, and circularly polarized to ellipticity angle $\chi$; see Fig. \ref{fig:fig1}(a). We use the standard definition for ellipticity angle $\chi = \arctan\left(b/a\right)$ where $a$ and $b$ are the major and minor axes of the polarization ellipse, $a \geq b$. The $1^{\text{st}}$-order output beam is used for trapping, but we note a residual, undeflected $0^{\text{th}}$-order beam and a weak negative $1^{\text{st}}$-order beam deflected by $-\theta$.

The MS device transforms the phase of the input beam according to the equation

\begin{eqnarray}
    \phi(x,y) = k_0 x \sin\left(\theta\right) + k_0   \left(\lvert f_1 \rvert  - \sqrt{x^2 + y^2 +f_1^2}  \right)  \nonumber 
    \\
    + k_0   \left(\sqrt{x^2 + y^2 +f_2^2} -\lvert f_2 \rvert \right)
    \label{eq:MSphase}
\end{eqnarray}

\noindent where $k_0 =\frac{2\pi}{\lambda}$ and $\lambda$ is the design wavelength.  We implement the phase map given by Eq. \ref{fig:fig1} by wrapping the phase between 0 and $2\pi$. The first term in Eq. \ref{eq:MSphase} represents grating-like deflection at our designed angle of $\theta = \pi/4$ and the last two terms represent a superposition of focusing lenses with focal length $f_i$ that controls the beam shape.  $f_1$ is a convex lens used to collimate the input mode from the fiber and $f_2$ is a concave lens that controls the beam divergence angle $\alpha$.  The birefringent properties of our MS originate in the designed asymmetry in the nanopillars' transverse dimensions, $L_x$ and $L_y$; refer to the scanning electron microscope (SEM) image in Fig. \ref{fig:fig1}(b). 
To obtain beams with circular polarization, we choose a subset of pillars that results in a fixed quarter-wave phase lag between the two orthogonal components of the electric field, analogous to the fast and slow axes in a QWP.  This amounts to computing the phase shifts for light linearly polarized along $\hat{x}$ ($\phi_x$) and $\hat{y}$ ($\phi_y$) and determining which pillars satisfy $\phi_x =\phi_y + \pi/2$. Our MSs produce left circular polarization (LCP) when the input field is polarized along an azimuthal angle between $L_x$ and $L_y$ of $\beta_{\text{in}} = \pi/4$ and right circular polarization (RCP) for $\beta_{\text{in}} = -\pi/4$ where $\beta_{\text{in}}$ is measured from $L_x$.

Our MSs are composed of TiO$_2$ nanopillars arranged on a grid of periodicity $p$ = 220 nm (330 nm) for wavelength 461 nm (689 nm) with ($L_x,L_y$) varying from $0.3p$ to $0.8p$; an SEM image of a sample MS is shown in Fig. \ref{fig:fig1}(b). We chose a pillar height of 900 nm to allow for efficient MS designs at each cooling wavelength to be integrated on single, 0.5-mm-thick fused silica wafer.  The nanostructures are fabricated by use of a Damascene process \cite{Fan2020IndependentMetasurfaces}, which allows for the definition of high-aspect-ratio features.  We use electron beam lithography (EBL) to form 900-nm deep holes in a resist layer which is subsequently filled with atomic layer deposition of TiO$_2$.  The film is planarized with an etch step and the resist layer is removed with lift-off, leaving isolated TiO$_2$ pillars.

Our trapping beam geometry resembles the conventional MOT shown in Fig. \ref{fig:fig1}(c) but is modified to enable the generation of three beams for each wavelength from a single wafer and all 12 beams from two wafers; see Fig. \ref{fig:fig1}(d). We create counter-propagating beams that are circularly polarized with the same handedness (LCP or RCP) relative to fixed lab coordinates to appropriately drive the $\sigma^{\pm}$ transitions see loop arrows) in the atoms and direct the atoms toward the center of the trap. The diameter, $d$, and power $P_{\text{MOT}}$, of the beams are chosen to give the desired trapping volume and to efficiently cool atoms from our thermal oven source based on the saturation intensity for each transition as well as the available fiber-coupled laser power, $P_{\text{fiber}}$. Our system is designed to achieve a total beam intensity of 33 mW/cm$^2$ and 42 mW/cm$^2$ for the 461 nm and 689 nm transitions respectively.

\begin{figure*}[t!]
    \centering
        \includegraphics[width=0.95\linewidth]{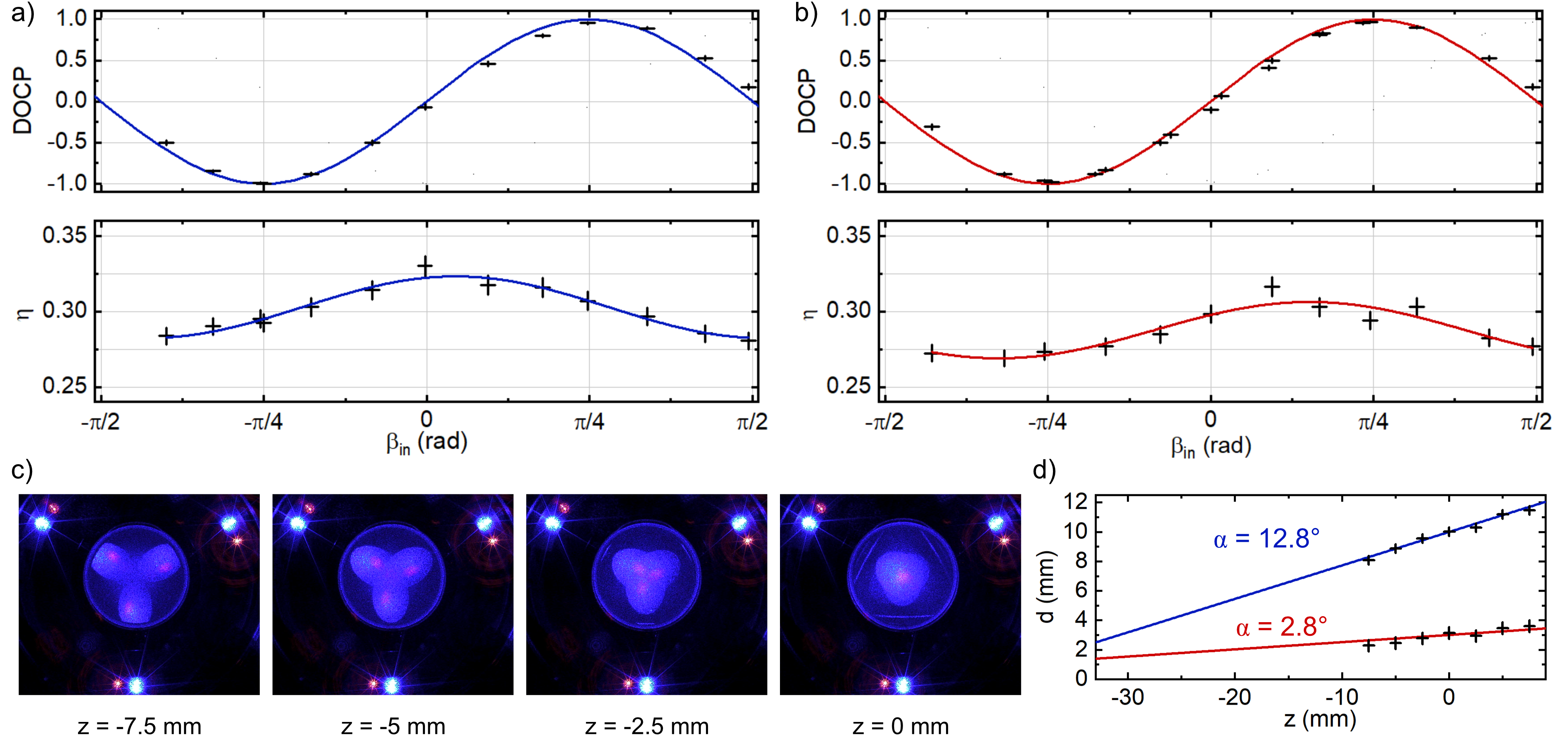}
\caption{
Characterization of MS beams at 461 nm and 689 nm.  (a) and (b) show the degree of circular polarization, $DOCP$, (top panel) and deflection efficiency, $\eta$, (bottom panel) as a function of input linear polarization angle, $\beta_{\text{in}}$, for 461 nm and 689 nm, respectively. (c) Images of the 461 nm (blue) and 689 nm (red) beams at various heights $z$ showing a common intersection at $z=0$. (d) Measured beam widths vs $z$ and comparison to the expected trend of our design values for $\alpha = 10.3^{\circ}$ (461 nm) and $\alpha=2.6^{\circ}$ (689 nm).  The error bars in (a-d) are instrumental uncertainty at $95\%$ confidence.  
}  
\label{fig:fig2}
\end{figure*}

Figure \ref{fig:fig2} presents our characterization of the 461-nm and 689-nm MSs for the three functionalities outlined in Fig. \ref{fig:fig1}(a). We first analyze the polarization transformation of the $1^{\text{st}}$-order beam through measurements of the degree of circular polarization, $DOCP$, as a function of the input azimuthal polarization angle, $\beta_{\text{in}}$; see the top panels of Figs. \ref{fig:fig2}(a) and \ref{fig:fig2}(b) respectively.  $DOCP$ represents the fraction of the optical power that is circularly polarized and is defined using the third Stokes parameter ($s_3$) so that $DOCP = \sin 2\chi$.  $DOCP = -1$ corresponds to LCP and $DOCP = +1$ to RCP.  Measurements at 461 nm and 689 nm are shown in the top panels of Fig. \ref{fig:fig2}(a) and \ref{fig:fig2}(b) respectively; the error bars are instrumental uncertainty at 95$\%$ confidence.  The equivalent QWP for our system should produce an output beam with an ellipticity angle $\chi_{\text{out}} = \beta_{\text{in}}$ from which we calculate the theoretical behavior $DOCP = \sin 2\beta_{\text{in}}$.  Our MSs reproduce this relation (blue and red lines) across the complete range of input angles, and reaches $|DOCP| \geq 0.95$ for both LCP and RCP at each wavelength.

Balanced optical powers in the counterpropagating beams are important for achieving balanced scattering forces and stable MOT operation.  We investigate the sensitivity of the beam power to misaligmnents in $\beta_{\text{in}}$ through measurements of the effective deflection efficiency defined as $\eta=P_{\text{MOT}}/P_{\text{fiber}}$.  We measure $\eta \approx 30 \%$ ($28.5 \%$) at 461 nm (689 nm), respectively, for both LCP and RCP and observe a weak sinusoidal dependence on input polarization angle; see bottom panels of Fig. \ref{fig:fig2}(a) and \ref{fig:fig2}(b) for data and fits.  We attribute the modulation to slightly different transmission for light polarized along the two transverse dimensions of each nanopillar.  We note that the overall efficiency is affected by a non-ideal overlap between the shape of the mode from the fiber and the shape of the finite-sized MS; see below for details about MS wafer layout.  When these effective aperture effects are taken into account, we find a fundamental MS efficiency of $\approx 45\%$ which is in good agreement with results from our simulations.  These measurements indicate that the dominant effect of misaligned polarization is to alter $DOCP$ and minimally affect $\eta$.

To reliably define the trapping volume and beam intensity, different MSs must generate beams with consistent orientations ($\theta = 45^{\circ}$) and shapes ($\alpha$).  We characterize these parameters by imaging the beams on a ground glass substrate at different heights above the wafer and performing routine image analysis to determine the beam positions and widths.  Figure \ref{fig:fig2}(c) shows images of the beams generated by a single wafer at various heights.  All six beams come to an intersection at the trap center located at the midpoint between the wafers; see far right panel for relative height $z=0$. Figure \ref{fig:fig2}(d) shows single beam measurements of the beam diameter with instrumental error bars.  At 461 nm (689 nm) we see excellent agreement with the expected trend shown as a blue line (red line) for the designed divergence angles of $\alpha = 10.3^{\circ}$ ($\alpha = 2.6^{\circ}$) and diameter at the trap center, $d=3$ mm ($d = 10$ mm).

\begin{figure}[ht!]
\centering
\includegraphics[width=\linewidth]{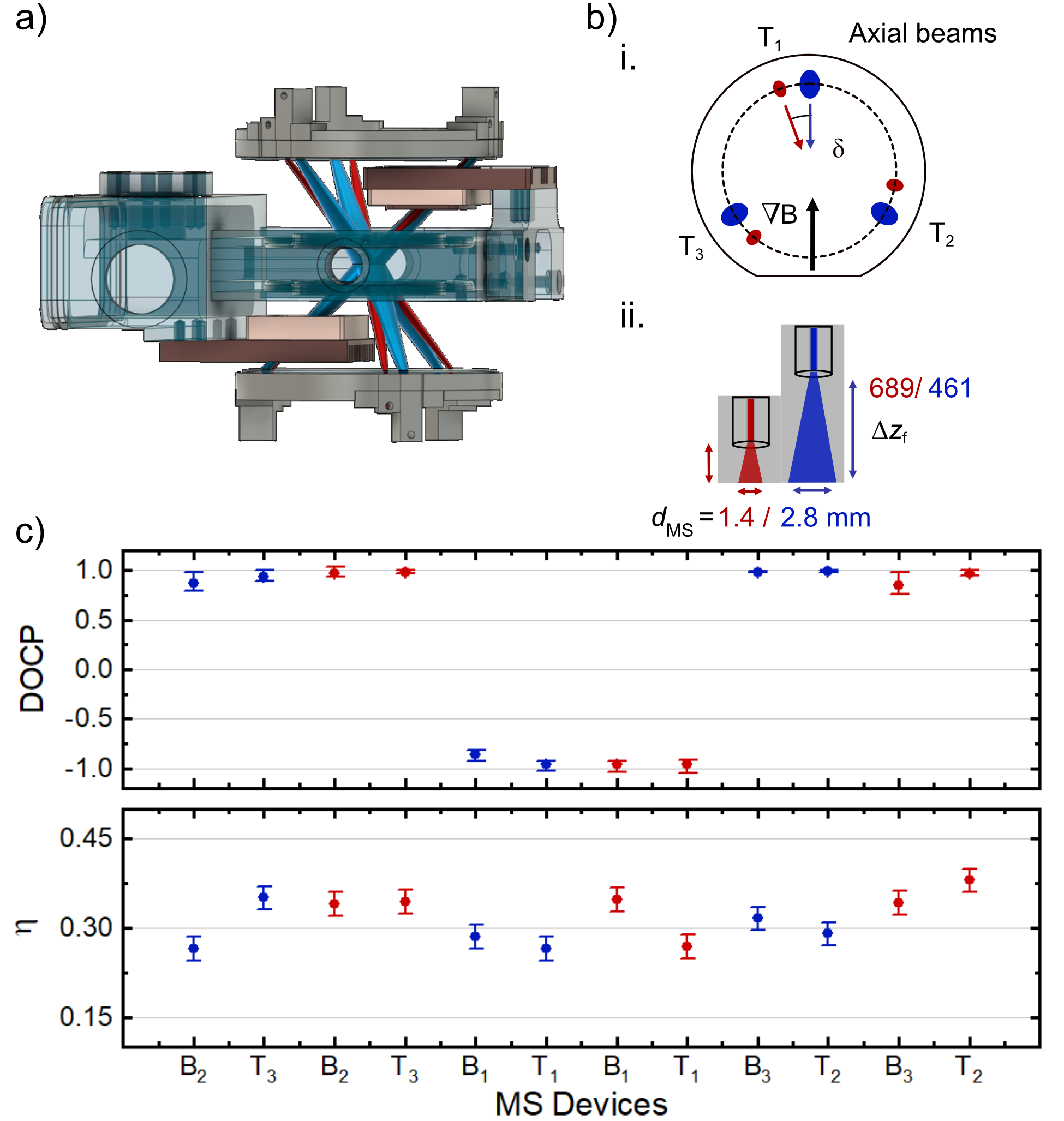}
\caption{ System-level integration.  (a) A CAD drawing showing two wafers mounted over our vacuum chamber and magnetic field coils.  (b)  Wafer-level integration. (b,i) The 461 nm (blue ovals) and 689 nm MSs (red ovals) are laid out on the wafer with a minimal separation $\delta$, ensuring similar geometry in relation to the magnetic field gradient $\nabla |\mathbf{B}|$.  (b,ii) Fiber fixtures that mechanically set the mode diameter at the MS, $d_{\text{MS}}$, by controlling the distance $\Delta z_{\text{f}}$. c)  $DOCP$ (top) and $\eta$ (bottom) measurements for all twelve metasurfaces used in our system. 
\label{fig:fig3}
}
\end{figure}

Figure \ref{fig:fig3} presents the system-level integration of our photonics with our compact vacuum chamber and magnetic field coils for a dual-color Sr MOT; see Fig. \ref{fig:fig3}(a) for a three-dimensional mechanical drawing of our system and the generated optical beams.  The 461-nm and 689-nm MSs are laid out on the wafer in a circle of radius $R$ with an azimuthal separation of $2\pi/3$; see Fig. \ref{fig:fig3} (b,i) for a diagram of the top wafer as viewed from trap center.  We use elliptically shaped MSs with semi-major ($A$) and -minor ($B$) radii related by $A=\sqrt{2} B$ to achieve a circular beam cross-section after being deflected by $\theta = \pi/4$.  We choose major diameters of $2A=2.8$ mm (1.4 mm) at 461 nm (689 nm), which is compatible with EBL write-times $< 8$ hours.  We label the MS devices on the top (bottom) wafer by $T_i$ ($B_i$) where $i=1,2,3$ is an index that describes the MS location counting clockwise starting with the device opposite the wafer flat. The $i=1$ devices generate the `axial beams' that pass through both magnetic field coils with a direction of propagation (blue and red arrows) that is approximately parallel to the direction of the strongest magnetic field gradient, $\nabla |\textbf{B}|$ (black arrow). The top and bottom wafers differ only in the choice of $\delta$ (see description below) and a mirror flip along the wafer midline parallel to the wafer flat.  This results in the desired counter-propagating beam geometry when both wafers are mounted with their flats oriented opposite each other.  The distance between the wafers is set with precision mounts to be $\Delta z = 2R + \epsilon$ where $\epsilon$ is a small correction that accounts for angular beam deviation in the vacuum chamber windows (2.54 mm thick, refractive index $n\approx 1.52$).

Our MS fiber coupling strategy robustly generates beams with the desired diameter at the trap center in a compact, plug-and-play package.  We offset the 689 nm and 461 nm devices by an angle of $\delta$ to allow for the inclusion of compact, LC fiber connectors.  Our choice of $\delta =  -10^\circ$ ($\delta =  +10^\circ$) for the top (bottom) wafer is sufficiently small to ensure that the 689-nm beams approximately maintain the geometry of the 461-nm beams with respect to $\nabla |\mathbf{B}|$. We note that the lateral offset between our magnetic field coils and non-cubic propagation angle $\theta = \pi/4$ means that neither wavelength experiences the traditional projections along $\nabla |\textbf{B}| $ for a cubic MOT.  We manually rotate the fiber ferrule into the keyless fixture to set the input polarization to generate LCP or RCP as desired.  This step represents the only manual alignment in our system but we note that a small alignment key could be included in future systems.  We achieve the desired trapping beam diameter through control of $f_i$ in the second term in Eq. \ref{eq:MSphase} and the $1/e^2$ diameter of the optical mode incident on the MS, $d_{\text{MS}}$.  The fibers that deliver 461-nm (PM-S405-XP) and 689-nm (PM630-HP) light have similar NAs and divergence angles so that $d_{\text{MS}}$ is primarily controlled by the distance between the fiber and MS, $\Delta z_{\text{f}}$. To minimize aperture losses due to finite MS size, we set $d_{\text{MS}} = 2A$ by using precision mechanical mounts to define $\Delta z_{\text{f}}$ to be 16 mm at 461 nm and 7.3 mm at 689 nm; see Fig. \ref{fig:fig3}(b,ii).  We also tightly control the lateral alignment between the optical fiber and the MS using high tolerance mechanical components to prevent errors in the beam-pointing direction that arise from off-axis coupling to the MSs.  These effects are most consequential for small-diameter beams where a small shift in position can lead to a large reduction in fractional overlap at the intersection volume.  For the 689-nm beams with a diameter of 3 mm at the trap center, a 140-$\mu$m error in the position of the fiber core can lead to a 50$\%$ reduction in the overlap at the trap center for two counterpropagating beams.

Finally, we evaluate the system-level performance of our MSs when used with COTS fiber components.  Figure \ref{fig:fig3}(c) shows measurements of the average $(DOCP,\eta)$ that can be obtained over reasonable misalignments of the input polarization and the error bars represent instrumental uncertainty at $95\%$ confidence.  MS devices at 461 nm (689 nm) are grouped according to pairs of counterpropagating beams (e.g., $B_2$ and $T_3$) and displayed as blue (red) circles.  The $i=1$ axial beams are set to LCP ($DOCP = -1$) and the $i=2,3$ beams to RCP ($DOCP = +1$).  Some of the devices shown have lower polarization purity than the $|DOCP| \geq 0.95$ level observed in Fig. \ref{fig:fig2}.   This reduction in $|DOCP|$ is caused by non-ideal performance of our visible PM fibers which are especially susceptible to misalignments of PM stress rods when using FC/APC connections. This misalignment results in a small degree of elliptical polarization at the output of the fiber (input to MS) that cannot be suppressed by azimuthal rotations of the ferrule to set $\beta_{\text{in}}$. Nevertheless, we find $|DOCP| \geq 0.85$ for devices in our system, which is sufficient for trapping \cite{Ferdinand2023TowardsPhotonics}.  As expected, $\eta$ is minimally affected by improper input polarization and we obtain an average efficiency across all beams of $\eta_{\text{avg}} = 30 \pm 3\%$ at 461 nm and $33 \pm 4\%$ at 689 nm (bottom panel); here the uncertainty range represents the statistical uncertainty at 65 $\%$ confidence.

In conclusion, our work demonstrates the successful multi-wavelength integration of six mm-scale metasurface optics on a single substrate to create a complex configuration of laser beams for optical cooling of Sr at 461 nm and 689 nm.  Our multifunction metasurface optics achieve precise control over beam pointing, divergence and polarization state without the need for bulk free-space optics. Our fiber-coupled approach streamlines the complexities of assembly and alignment through a foundry-compatible integration strategy and represents a significant step towards portable cold atom systems with applications in geodesy, inertial sensing and optical atomic clocks.

\begin{backmatter}
\bmsection{Funding} 
Defense Advanced Research Projects Agency(DARPA) (A-PhI);  National Institute of Standards
and Technology (NIST)

\end{backmatter}

\bibliography{references}

\end{document}